\documentclass{PoS}
\usepackage{graphicx}
\usepackage{verbatim}
\usepackage{amsmath}
\usepackage{epsfig}

\title{Azimuthal angle decorrelation of Mueller--Navelet jets at NLO}
\ShortTitle{Azimuthal angle decorrelation of Mueller--Navelet jets at NLO}

\author{\speaker{A. Sabio Vera}\\
        Physics Department, Theory Division, CERN, CH-1211 Geneva 23, Switzerland\\
        E-mail: \email{Agustin.Sabio.Vera@cern.ch}}

\author{F. Schwennsen\\
        II. Institut f\"ur Theoretische Physik, Universit\"at Hamburg, Luruper Chaussee 149, D-22761~Hamburg, Germany\\
        E-mail: \email{florian.schwennsen@desy.de}}

\abstract{\noindent
In this contribution we study azimuthal angle decorrelation in inclusive 
dijet cross sections taking into account the next--to--leading (NLO) 
corrections 
to the BFKL kernel while keeping the jet vertices at leading order. 
We show how the angular decorrelation for jets with a wide relative 
separation in rapidity largely decreases when the NLO corrections are 
included.}

\FullConference{DIFFRACTION 2006 - International Workshop on Diffraction in High-Energy Physics \\
                 September 5-10 2006\\
                 Adamantas, Milos island, Greece}

\begin{document}
\newcommand{\non}{\nonumber\\}
\newcommand{\asbar}{\bar{\alpha}_s}
\newcommand{\plusinu}[1]{\left(#1\right)^{i\nu-\frac{1}{2}}}
\newcommand{\minusinu}[1]{\left(#1\right)^{-i\nu-\frac{1}{2}}}
\newcommand{\plusinup}[1]{\left(#1\right)^{i\nu'-\frac{1}{2}}}
\newcommand{\minusinup}[1]{\left(#1\right)^{-i\nu'-\frac{1}{2}}}

\section{Introduction}

One of the important questions still open in Quantum Chromodynamics 
is how to describe scattering amplitudes in the Regge limit where 
the center--of--mass energy, $s$, is much larger than all other 
Mandelstam invariants and mass scales. In this region it is needed to resum 
logarithmically enhanced contributions of the form 
$\left(\alpha_s \ln{s}\right)^n$ to all orders. This is achieved using the 
leading--logarithmic (LL) Balistky--Fadin--Kuraev--Lipatov (BFKL) evolution 
equation~\cite{FKL}.

Observables where BFKL effects should be dominant require of a large  
energy to build up the parton evolution and the presence of two large and 
similar transverse scales. An example is the inclusive hadroproduction of two 
jets with  large and similar transverse momenta and a big relative separation 
in rapidity, the so--called Mueller--Navelet jets. When Y, the distance in 
rapidity between the most forward and backward jets, is not large a fixed 
order perturbative analysis should be enough to describe the cross section but 
when it increases a BFKL resummation of $\left(\alpha_s {\rm Y}\right)^n$ 
terms is needed. This observable was proposed in Ref.~\cite{Mueller:1986ey} as 
a clean configuration to look for BFKL effects at hadron colliders. 
A power--like rise for the partonic cross section was predicted. At a more 
exclusive level one can study  the azimuthal angle 
decorrelation of the pair of jets. 
BFKL enhances soft real emission as Y increases reducing the angular correlation in transverse plane originally present in the back--to--back Born configuration. 
The LO prediction for this azimuthal dependence was first investigated in 
Ref.~\cite{LLResults} where it was shown that it overestimates the rate of 
decorrelation when compared with the Tevatron data~\cite{Tevatron}.  
In the present contribution we summarize the work of Ref.~\cite{Vera:2006un}
where $\alpha_s \left(\alpha_s {\rm Y}\right)^n$ 
next--to--leading logarithmic (NLL) corrections to the BFKL 
kernel~\cite{FLCC} were included. Previous numerical studies of this 
kernel were performed in Ref.~\cite{Andersen:2003wy} using an implementation 
of the NLL iterative solution proposed in Ref.~\cite{Andersen:2003an}. 
In coming works we would like to include the next--to--leading order (NLO) jet 
vertex~\cite{impactfactors}, investigate more convergent versions of the 
kernel~\cite{improvedkernel} and include parton distributions effects. 
It will be also interesting to study if the jet definition in the NLO kernel  
proposed in Ref.~\cite{Bartels:2006hg} has sizable phenomenological 
implications.  
We believe Mueller--Navelet jets should be an important test of our 
understanding of small $x$ resummation to be performed at the Large Hadron 
Collider at CERN.

\section{The dijet partonic cross section}

We are interested in the calculation of the partonic cross 
section parton + parton $\rightarrow$ jet + jet + soft emission, with the two jets having transverse momenta $\vec{q}_1$ and $\vec{q}_2$ and being produced 
at a large relative rapidity separation Y. In the particular case of 
gluon--gluon scattering we have
\begin{eqnarray}
\frac{d {\hat \sigma}}{d^2\vec{q}_1 d^2\vec{q}_2} &=& \frac{\pi^2 {\bar \alpha}_s^2}{2} 
\frac{f \left(\vec{q}_1,\vec{q}_2,{\rm Y}\right)}{q_1^2 q_2^2},
\end{eqnarray}
with ${\bar \alpha}_s = \alpha_s N_c/\pi$ and $f$ the gluon Green's 
function which is the solution to the BFKL equation.
 
The partonic cross section is obtained by integration over the 
phase space of the two emitted gluons together with the jet vertices:
\begin{eqnarray}
{\hat \sigma} \left(\alpha_s, {\rm Y},p^2_{1,2}\right) &=&
\int d^2{\vec{q}_1} \int d^2{\vec{q}_2} \,
\Phi_{\rm jet_1}\left(\vec{q}_1,p_1^2\right)
\,\Phi_{\rm jet_2}\left(\vec{q}_2,p_2^2\right)\frac{d {\hat \sigma}}{d^2\vec{q}_1 d^2\vec{q}_2}.
\end{eqnarray}
For the jet vertices we use the LO ones $\Phi_{\rm jet_i}^{(0)} \left(\vec{q},p_i^2\right) = \theta \left(q^2-p_i^2\right)$, where $p_i^2$ corresponds to a resolution scale for the transverse momentum of 
the gluon jet. In this way a full NLO accuracy is not 
achieved but it is possible to pin down those effects stemming 
from  the gluon Green's function. To extend this analysis it would be needed to calculate the 
Mellin transform of the NLO jet vertices in Ref.~\cite{impactfactors} where 
the definition of a jet is much more involved than 
here. To proceed further it is very convenient to recall the work of 
Ref.~\cite{Ivanov:2005gn} and use the operator representation 
${\hat q} \left|\vec{q}_i\right> = \vec{q}_i \left|\vec{q}_i\right>$ 
with normalization $\left<\vec{q}_1\right|{\hat 1}\left|\vec{q}_2\right> = 
\delta^{(2)} \left(\vec{q}_1-\vec{q}_2\right)$. In this notation the BFKL 
equation simply reads $\left(\omega - {\hat K}\right) {\hat f}_\omega = {\hat 1} $ where we have performed a Mellin transform in rapidity space:
\begin{eqnarray}
f\left(\vec{q}_1,\vec{q}_2,{\rm Y}\right) &=& \int \frac{d\omega}{2 \pi i}
e^{\omega {\rm Y}} f_\omega \left(\vec{q}_1,\vec{q}_2\right).
\end{eqnarray}
The kernel has the  expansion ${\hat K}= {\bar \alpha}_s {\hat K}_0 + {\bar \alpha}_s^2 {\hat K}_1 + \dots $. To NLO accuracy this implies that the solution can be written as
\begin{eqnarray}
{\hat f}_\omega = \left(\omega - {\bar \alpha}_s {\hat K}_0\right)^{-1}
+ {\bar \alpha}_s^2 \left(\omega - {\bar \alpha}_s {\hat K}_0\right)^{-1} 
{\hat K}_1 \left(\omega - {\bar \alpha}_s {\hat K}_0\right)^{-1} + 
{\cal O}\left({\bar \alpha}_s^3\right). 
\label{opGGF}
\end{eqnarray}
The basis on which to express the cross section reads 
\begin{eqnarray}
\left< \vec{q}\right|\left.\nu,n\right> &=& \frac{1}{\pi \sqrt{2}} 
\left(q^2\right)^{i \nu -\frac{1}{2}} \, e^{i n \theta}. 
\label{eignfns}
\end{eqnarray}
The action of the NLO kernel on this basis, which was calculated in 
Ref.~\cite{Kotikov:2000pm}, contains non--diagonal terms and can be written as 
\begin{eqnarray}
{\hat K} \left|\nu,n\right> &=& \left\{\frac{}{}{\bar \alpha}_s \, \chi_0\left(\left|n\right|,\nu\right) + {\bar \alpha}_s^2 \, \chi_1\left(\left|n\right|,\nu\right) \right.\nonumber\\
&&\left.\hspace{-2cm}+\,{\bar \alpha}_s^2 \,\frac{\beta_0}{8 N_c}\left[2\,\chi_0\left(\left|n\right|,\nu\right) \left(i \frac{\partial}{\partial \nu}+ \log{\mu^2}\right)+\left(i\frac{\partial}{\partial \nu}\chi_0\left(\left|n\right|,\nu\right)\right)\right]\right\} \left|\nu,n\right>,
\label{opKernel}
\end{eqnarray}
where, from now on, ${\bar \alpha}_s$ stands for ${\bar \alpha}_s \left(\mu^2\right)$, the coupling evaluated at the renormalization point $\mu$ in the $\overline{\rm MS}$ scheme. The first line of Eq.~(\ref{opKernel}) corresponds 
to the scale invariant sector of the kernel. Both functions $\chi_0$ and 
$\chi_1$ can be found in Ref.~\cite{Vera:2006un}.

\begin{figure}[tbp]
\centering
  \epsfig{width=8cm,file=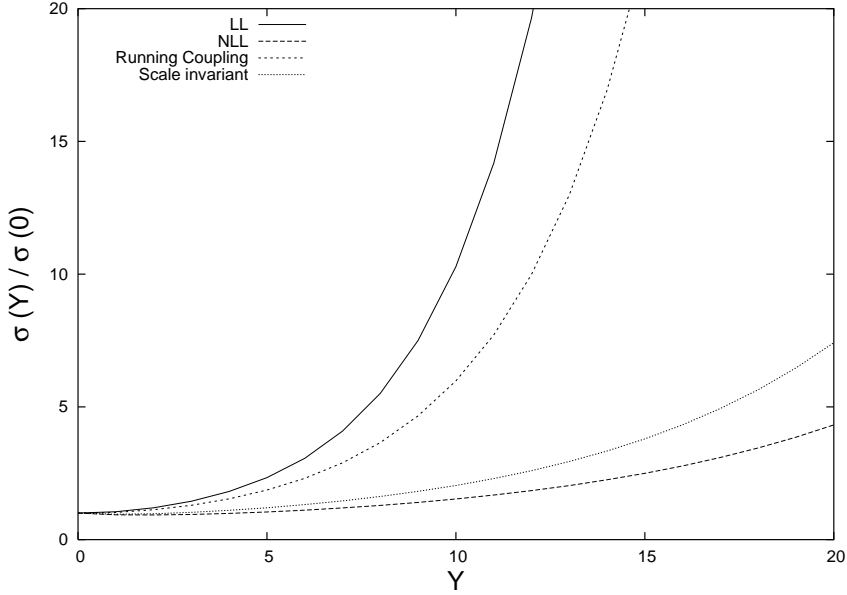,angle=-90}
\caption{Evolution of the partonic cross section with the rapidity separation of the dijets. }
\label{SectionconY}
\end{figure}

\begin{figure}[tbp]
\centering
\hspace{0.4cm}
\epsfig{width=6cm,file=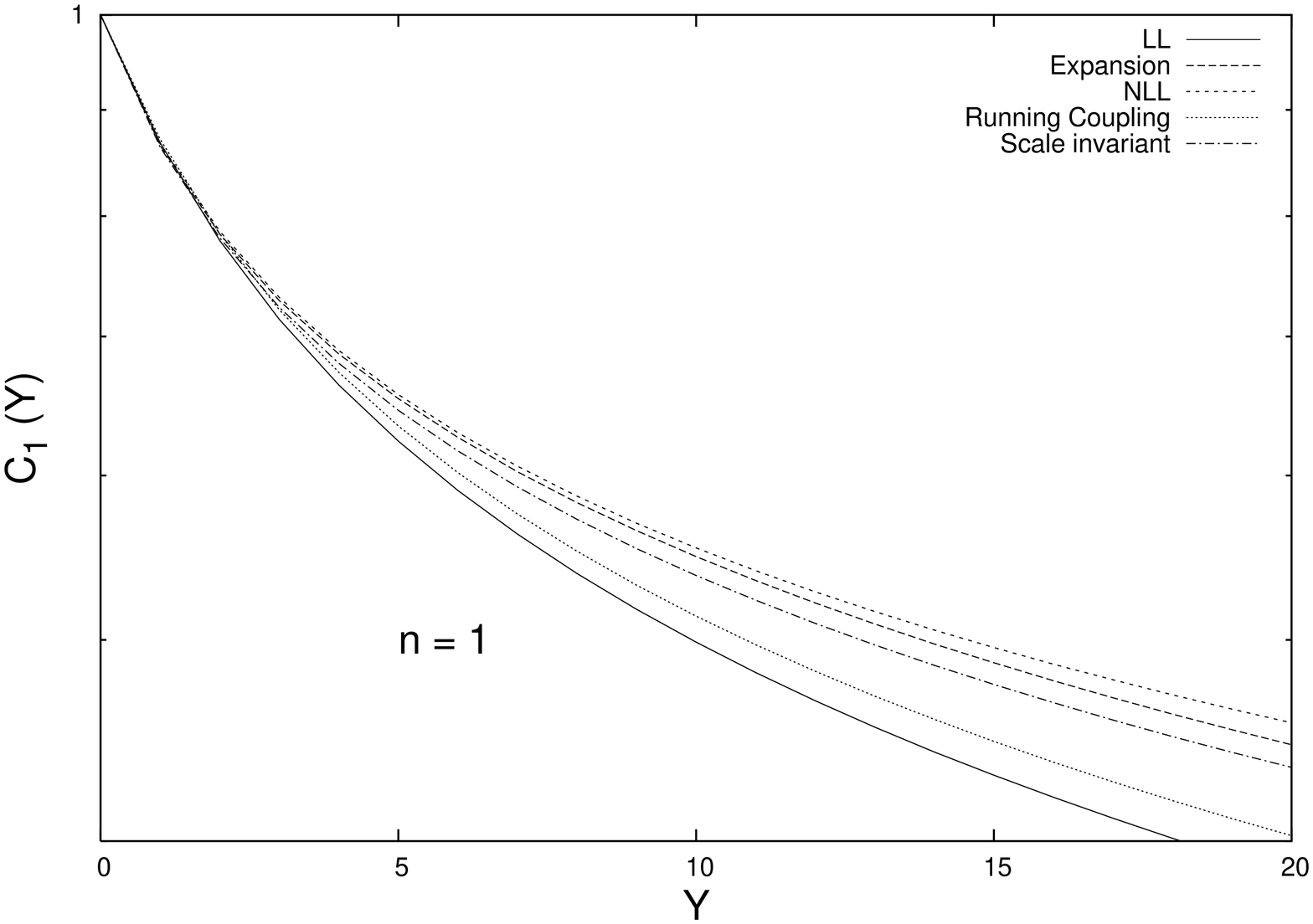,angle=-90}
\epsfig{width=6cm,file=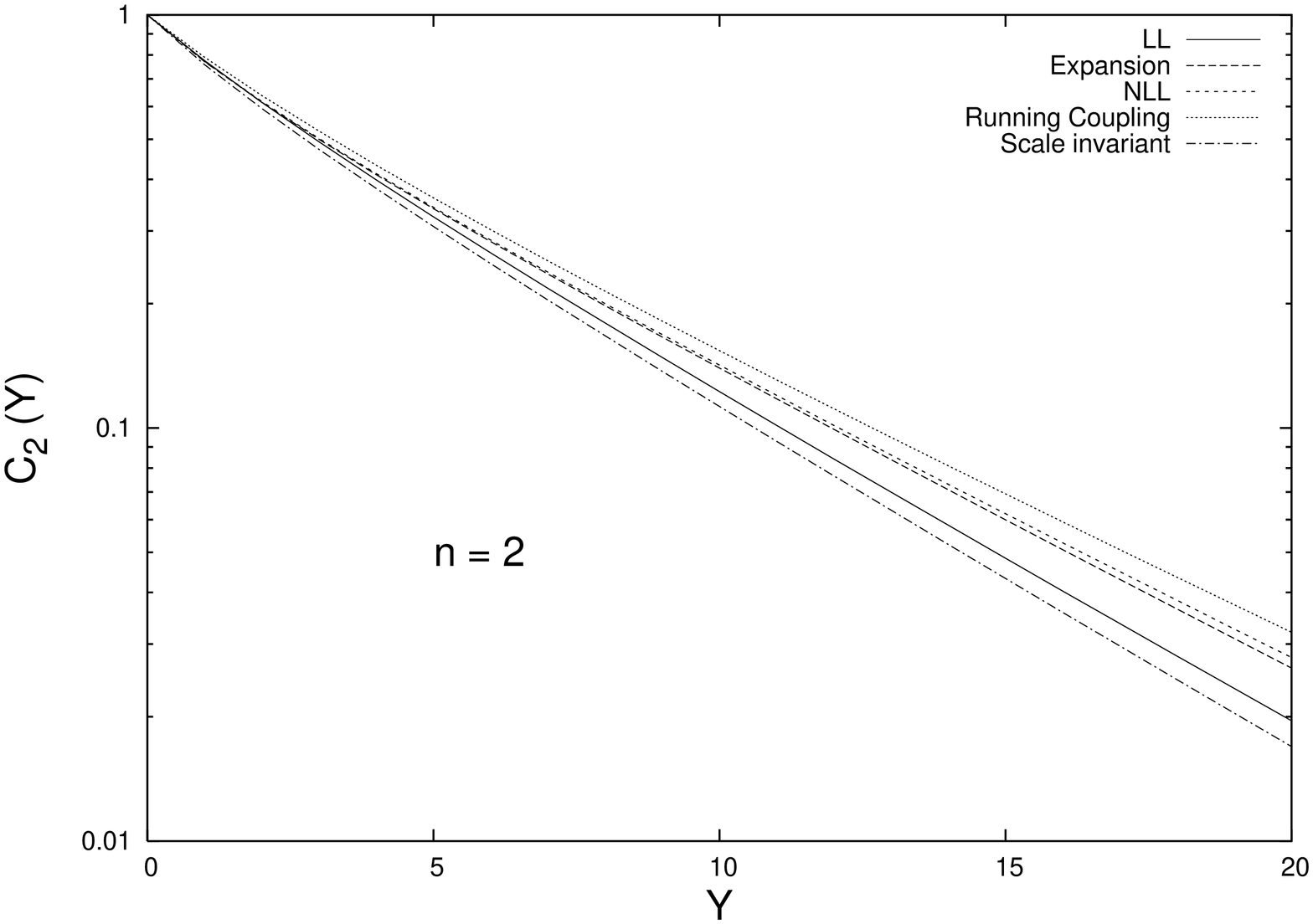,angle=-90}
\caption{Evolution in Y of the ${\cal C}_n ({\rm Y})$ coefficients for $n=1,2,3$.}
\label{CnvsY123}
\end{figure}

\begin{figure}[tbp]
\centering
  \epsfig{width=8cm,file=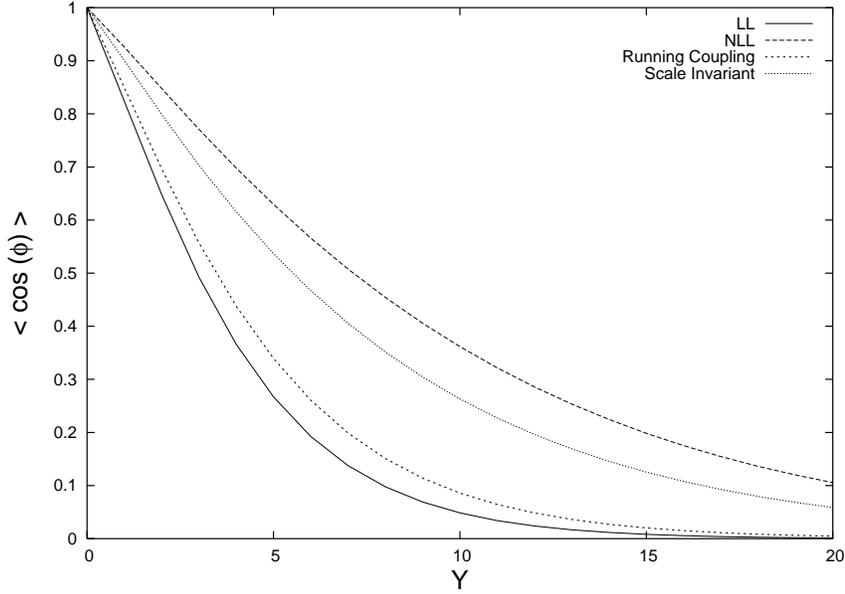 ,angle=-90}
\caption{Dijet azimuthal angle decorrelation as a function of their separation in rapidity.}
\label{Cos1Y}
\end{figure}

\begin{figure}[tbp]
\centering
  \epsfig{width=8cm,file=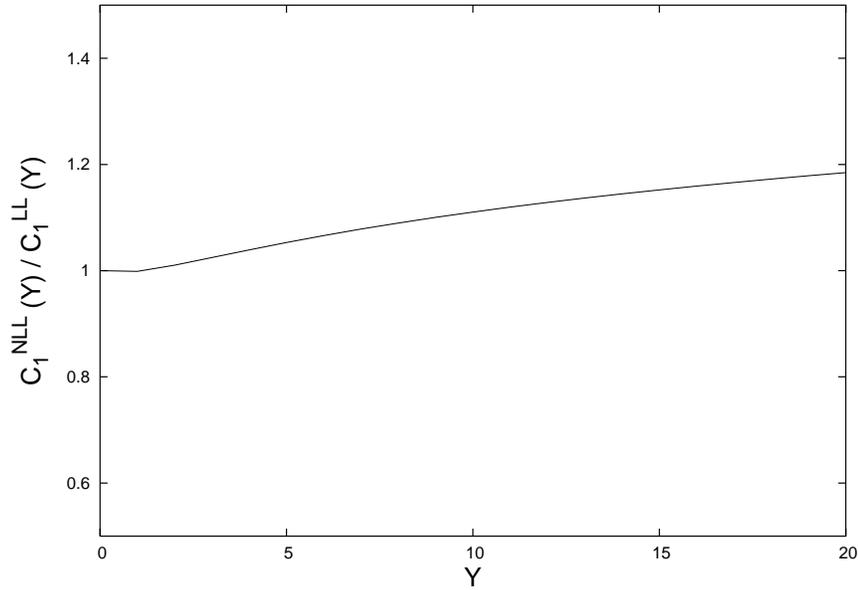 ,angle=-90}
\caption{Comparative ratio between the NLL and LL coefficients for $n=1$ conformal 
spin.}
\label{CosRatio}
\end{figure}

Using the notation
\begin{eqnarray}
c_{1}\left(\nu\right) &\equiv& \frac{1}{\sqrt{2}}\frac{1}{\left(\frac{1}{2}-i \nu\right)}\left(p_1^2\right)^{i \nu - \frac{1}{2}} ,
\label{IFproj}
\end{eqnarray}
and $c_2$ being the complex conjugate of this expression with $p_1^2$ replaced 
by $p_2^2$, we can then write the cross section as
\begin{eqnarray}
{\hat \sigma} \left(\alpha_s, {\rm Y},p_{1,2}^2\right) &=&
\frac{\pi^2 {\bar \alpha}_s^2}{2} \sum_{n=-\infty}^\infty \int_{-\infty}^\infty d \nu 
\,e^{{\bar \alpha}_s \chi_0\left(\left|n\right|,\nu\right) {\rm Y}} 
c_1\left(\nu\right) c_2\left(\nu\right) \delta_{n,0} \label{logdercross}\\
&&\hspace{-3.3cm}\times \left\{1+{\bar \alpha}_s^2 \, {\rm Y} \left[\chi_1\left(\left|n\right|,\nu\right)+\frac{\beta_0}{4 N_c} \left(\log{(\mu^2)}+ \frac{i}{2} \frac{\partial}{\partial \nu}\log{\left(\frac{c_1\left(\nu\right)}{c_2\left(\nu\right)}\right)}+ \frac{i}{2} \frac{\partial}{\partial \nu}\right)\chi_0\left(\left|n\right|,\nu\right)\right]\right\}.\nonumber
\end{eqnarray}
The angular differential cross section in the case where the two resolution 
momenta are equal, $p_1^2 = p_2^2 \equiv p^2$, and using the notation 
$\phi = \theta_1-\theta_2 - \pi$  can be expressed as
\begin{eqnarray}
\frac{d {\hat \sigma}\left(\alpha_s, {\rm Y},p^2\right)}{d \phi}  &=&
\frac{\pi^2 {\bar \alpha}_s^2}{4 p^2} \sum_{n=-\infty}^\infty 
\frac{1}{2 \pi}e^{i n \phi} \int_{-\infty}^\infty d \nu 
\,e^{{\bar \alpha}_s \chi_0\left(\left|n\right|,\nu\right) {\rm Y}} 
\frac{1}{\left(\frac{1}{4}+\nu^2\right)}\nonumber\\
&&\hspace{-3cm}\times \left\{1+{\bar \alpha}_s^2 \, {\rm Y} \left[\chi_1\left(\left|n\right|,\nu\right)-\frac{\beta_0}{8 N_c} \chi_0\left(\left|n\right|,\nu\right) \left(2 \log{\left(\frac{p^2}{\mu^2}\right)}+\frac{1}{\left(\frac{1}{4}+\nu^2\right)}\right)\right]\right\}.
\label{Noexp}
\end{eqnarray}
It can be conveniently rewritten as
\begin{eqnarray}
\frac{d {\hat \sigma}\left(\alpha_s, {\rm Y},p^2\right)}{d \phi}  &=&
\frac{\pi^3 {\bar \alpha}_s^2}{2 p^2} \frac{1}{2 \pi}\sum_{n=-\infty}^\infty 
e^{i n \phi} {\cal C}_n \left({\rm Y}\right),
\end{eqnarray}
with
\begin{eqnarray}
{\cal C}_n \left({\rm Y}\right) =
\int_{-\infty}^\infty \frac{d \nu}{2 \pi}\frac{e^{{\bar \alpha}_s \left(p^2\right){\rm Y} \left(\chi_0\left(\left|n\right|,\nu\right)+{\bar \alpha}_s  \left(p^2\right) \left(\chi_1\left(\left|n\right|,\nu\right)-\frac{\beta_0}{8 N_c} \frac{\chi_0\left(\left|n\right|,\nu\right)}{\left(\frac{1}{4}+\nu^2\right)}\right)\right)}}{\left(\frac{1}{4}+\nu^2\right)}.
\label{Cn}
\end{eqnarray}
The coefficient governing the energy dependence of the cross section 
corresponds to $n=0$:
\begin{eqnarray}
{\hat \sigma}\left(\alpha_s, {\rm Y},p^2\right) &=& 
\frac{\pi^3 {\bar \alpha}_s^2}{2 p^2} \, {\cal C}_0 \left({\rm Y}\right).
\end{eqnarray}
The rise in rapidity of this observable, with $p = 30 \, {\rm GeV}$, 
$n_f = 4$ and $\Lambda_{\rm QCD} = 0.1416$ GeV, is shown in 
Fig.~\ref{SectionconY}. Clearly the NLL intercept is very much reduced with 
respect to the LL case. In our plots we show the results for a LO version 
of the kernel, which has the largest intercept. Then we plot the effects of 
the running of the coupling alone, which tends to reduce the growth with 
rapidity. And finally we show the contribution from the scale invariant 
pieces which provide the most negative pieces.
The remaining coefficients with $n \geq 1$ all decrease with Y. This can be 
seen in the plots of Fig.~\ref{CnvsY123}.
The consequence of this decrease is that the angular correlations also 
diminish as the rapidity interval between the jets gets larger. This point 
can be studied in detail using the mean values 
\begin{eqnarray}
\left<\cos{\left( m \phi \right)}\right> &=& \frac{{\cal C}_m \left({\rm Y}\right)}{{\cal C}_0\left({\rm Y}\right)}.
\end{eqnarray}
$\left<\cos{\left(\phi\right)}\right>$ is calculated in Fig.~\ref{Cos1Y}. The 
most important consequence of this plot is that the NLL effects dramatically 
decrease the azimuthal angle decorrelation. This is already the case when 
only the running of the coupling is introduced but the scale invariant terms 
make this effect much bigger. This is encouraging from the 
phenomenological point of view given that the data at the Tevatron typically 
have lower decorrelation than predicted by LL BFKL or LL with running coupling. 
It is worth noting that the 
difference in the prediction for decorrelation between 
LL and NLL is mostly driven by the $n=0$ conformal spin. This can be 
understood looking at the ratio
\begin{eqnarray}
\frac{\left<\cos{\left(\phi \right)}\right>^{\rm NLL}}{\left<\cos{\left(\phi \right)}\right>^{\rm LL}} &=& \frac{{\cal C}_1^{\rm NLL} \left({\rm Y}\right)}{{\cal C}_0^{\rm NLL}\left({\rm Y}\right)}\frac{{\cal C}_0^{\rm LL} \left({\rm Y}\right)}{{\cal C}_1^{\rm LL}\left({\rm Y}\right)},
\end{eqnarray}
and noticing that 
\begin{eqnarray}
1.2 > \frac{{\cal C}_1^{\rm NLL} \left({\rm Y}\right)}{{\cal C}_1^{\rm LL}\left({\rm Y}\right)} > 1.
\end{eqnarray}
This ratio is calculated in Fig.~\ref{CosRatio}.
This point is a consequence of the good convergence in terms of asymptotic 
intercepts of the NLL BFKL calculation for conformal spins larger than zero. 
In particular the $n=1$ case is special in that the property of zero intercept 
at LL, $\chi_0(1,1/2) = 0$, is preserved under radiative corrections since
\begin{eqnarray}
\chi_1\left(1,\frac{1}{2}\right) = 
{\cal S} \chi_0 \left(1, \frac{1}{2}\right)
+ \frac{3}{2} \zeta\left(3\right) 
- \frac{\beta_0}{8 N_c}\chi_0^2\left(1, \frac{1}{2}\right)
+\frac{\psi''\left(1\right)}{2} - \phi\left(1, \frac{1}{2}\right)
\end{eqnarray}
is also zero. 

\section{Conclusions}

An analytic procedure has been presented to calculate the effect of higher order 
corrections in the description of Mueller--Navelet 
jets where two jets with moderately high and similar transverse momentum
 are produced at a large 
relative rapidity separation in hadron--hadron collisions. This is a 
promising observable to study small $x$ physics at the Large Hadron Collider 
at CERN given its large energy range. The focus of 
the analysis has been on those effects with direct origin in the NLO BFKL kernel, 
while the jet vertices have been considered at LO accuracy. It 
has been shown how the growth with energy of the cross section is reduced 
when going from a LL to a NLL approximation, 
and how the azimuthal angle decorrelations largely decrease due to the 
higher order effects. The present study has been performed at partonic 
level while the implementation of a full analysis, including parton 
distribution 
functions, NLO jet vertices and the investigation of 
collinearly improved kernels, will be published elsewhere~\cite{FA}. 

\noindent
{\bf Acknowledgments:} 
F.S. is supported by the Graduiertenkolleg ``Zuk\"unftige Entwicklungen in der Teilchenphysik''. 
Discussions with J.~Bartels, D.~Dobur, L.~Lipatov and A.~Papa are acknowledged.

\end{document}